\documentclass[aps,prb,twocolumn,groupedaddress]{revtex4-1} 

\usepackage{epsfig}
\usepackage{array}
\newcommand{\PreserveBackslash}[1]{\let\temp=\\#1\let\\=\temp}
\newcolumntype{C}[1]{>{\PreserveBackslash\centering}p{#1}}
\newcolumntype{R}[1]{>{\PreserveBackslash\raggedleft}p{#1}}
\newcolumntype{L}[1]{>{\PreserveBackslash\raggedright}p{#1}}

\begin{document}


\title{A new structure of two-dimensional allotropes of group \uppercase\expandafter{\romannumeral5} elements}


\author{Ping Li}
\author{Weidong Luo}
\email{wdluo@sjtu.edu.cn}
\affiliation{Key Laboratory of Artificial Structures and Quantum Control, Department of Physics and Astronomy, Shanghai Jiao Tong University, Shanghai 200240, China\\Institute of Natural Sciences, Shanghai Jiao Tong University, Shanghai 200240, China\\Collaborative Innovation Center of Advanced Microstructures, Nanjing 210046, China}



\date{\today}

\begin{abstract}
The elemental two-dimensional (2D) materials such as graphene, silicene, germanene, and black phosphorus have attracted considerable attention due to their fascinating physical properties. Structurally they possess the honeycomb or distorted honeycomb lattices, which are composed of six-atom rings. Here we find a new structure of 2D allotropes of group \uppercase\expandafter{\romannumeral5} elements composed of eight-atom rings, which we name as the octagonal tiling (OT) structure. First-principles calculations indicate that these allotropes are dynamically stable and are also thermally stable at temperatures up to 600 K. These allotropes are semiconductors with band gaps ranging from 0.3 to 2.0 eV, thus they are potentially useful in near- and mid-infrared optoelectronic devices. OT-Bi is also a 2D topological insulator (TI) with a band gap of 0.33 eV, which is the largest among the reported elemental 2D TIs, and this gap can be increased further by applying compressive strains.
\end{abstract}


\maketitle

The successful exfoliation of graphene from graphite brings two-dimensional (2D) materials from the theoretical domain to the real world \cite{p1,p2,p3}. Since then, many more 2D materials have been studied. For example, graphene, in a honeycomb structure, has attracted tremendous attention due to its excellent dynamical properties and unique electronic structures \cite{p3,p4,p5,p6}. Silicene \cite{p7,p8}, germanene \cite{p7, p9,p10}, stanene \cite{p11,p12} and Bi(111) bilayer \cite{p13,p14,p15,p16,p17,p18,p19} are analogues of graphene and they have the buckled honeycomb structures. Recent studies \cite{p7,p11,p13,p14} have proposed that these analogues are 2D topological insulators (TIs), also called quantum spin Hall (QSH) insulators, which are characterized by coexistence of gapped bulk states and gapless edge states protected by time reversal symmetry \cite{p20,p21}. They have attracted much attention due to fundamental interest and potential applications in spintronics and topological quantum computations \cite{p20,p21}. The nontrivial energy gap of Bi(111) bilayer is about 0.2 eV \cite{p14,p15}, which is the largest among the reported elemental  2D TIs until now. This is important, because for room temperature utilizations, energy gaps of TIs must be large enough to suppress the effects of thermal perturbations on transport properties. Encouraging signatures have been experimentally observed in this material, such as the existence of edge states \cite{p18,p19}. Black phosphorus (BP) layers have been investigated for their potentials for optoelectronics and electronics applications. These layers have the puckered honeycomb structure and can be exfoliated from the bulk materials \cite{p23}. They exhibit high-mobility transport anisotropy and linear dichroism \cite{p23,p24}, and the energy gaps can be tuned from about 0.3 to 1.5 eV by varying the layer thickness \cite{p22}. Recently, BP field-effect transistors have been experimentally fabricated \cite{p25}.

The 2D elemental materials discussed above all have the honeycomb or distorted honeycomb structures. In common, they are composed of six-atom rings. Carbon has been reported to have many other 2D allotropes \cite{p26,p27}, while the known 2D allotropes for other elements are quite limited. A pseudocubic \{012\}-oriented allotrope of Bi can be realized on Si(111) surface \cite{p28}, and it was proposed to be in the BP-like puckered-layer structure. In three dimensions, the group \uppercase\expandafter{\romannumeral5} elements show rich allotropic transformation because their semi-metallic bonding character can be easily shifted to either metallic or covalent side. Different allotropes can have very different physical properties. For example, As and Sb can exist in metallic forms or as semiconductors of different band gaps, depending on their crystal structures \cite{p31}. For two dimensions, it is also of interests to explore new allotropes of the group \uppercase\expandafter{\romannumeral5} elements and investigate their electronic properties.

In this article, we report a new structure of 2D allotropes of the group \uppercase\expandafter{\romannumeral5} elements (P, As, Sb and Bi) composed of eight-atom rings. These allotropes contain unique atomic octagonal tiling (OT) patterns and are shown to be stable by using first-principles calculations. All these OT ultra-thin films are semiconductors with band gaps ranging from 0.3 to 2.0 eV, covering the entire gap range of BP and transition-metal dichalcogenides (TMDCs), thus they are candidates for various near- and mid-infrared optoelectronic applications. In addition, the OT-Bi bilayer is a 2D TI with a band gap of 0.33 eV, which is larger than the gap of the Bi(111) bilayer and is the largest among reported elemental 2D TIs.

\vspace{5mm}
\textbf{Result}

\begin{figure}[!htb]
\includegraphics [width=7.6cm]{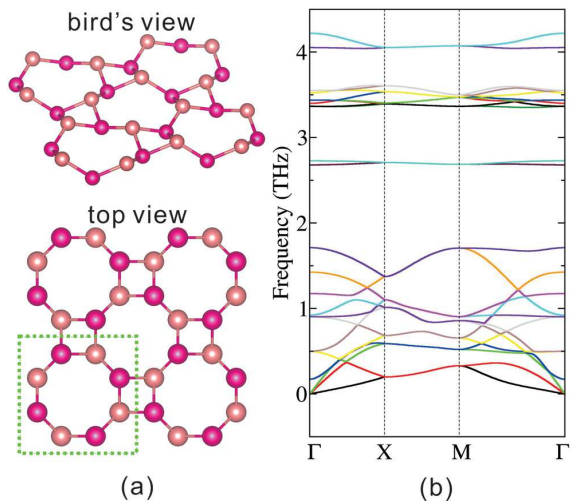}
\caption{(a) Atomic structures of the 2D octagonal tiling structure. The two colours indicate atoms in different heights of the bilayer. (b) The calculated phonon spectrum of the OT-Bi bilayer. }
\label{figure1}
\end{figure}

The 2D lattice structure model studied in this work is shown in Fig.~\ref{figure1} (a), and its unit cell contains eight atoms. The atoms have two different heights, so it can be viewed as a bilayer structure. From top view, the atoms form a beautiful octagonal tiling, which is composed of octagons with alternating long and short edges and smaller squares. In the following, we will show that all the OT-X (X = P, As, Sb and Bi) bilayers are metastable 2D allotropes of group \uppercase\expandafter{\romannumeral5} elements and they exhibit interesting structural and electronic properties.

Structural optimizations of the octagonal tiling structure were performed for the group \uppercase\expandafter{\romannumeral5} elements P, As, Sb and Bi. The optimized structural parameters such as the lattice constant \textit{a}, the bilayer thickness \textit{d}, and the bond lengths \textit{l}$_1$ and \textit{l}$_2$ of the long and short edges of the octagons are listed in Table~\ref{table1}. These parameters increase as the atomic size increases. Comparing with the reference structures such as the monolayer black P and As, bilayer buckled honeycomb Sb and Bi, the energies of these OT allotropes are slightly higher. Typically, the energy differences are about 70 to 80 meV per atom.

\begin{table}[tbp]
\caption{The calculated lattice constant \textit{a}, the bilayer thickness \textit{d}, and the bond lengths \textit{l}$_1$ and \textit{l}$_2$ of the long and short edges of the octagons of the OT allotropes (\textit{a}, \textit{d}, \textit{l}$_1$ and \textit{l}$_2$ all in units of angstrom).}
\centering  
\begin{tabular}{L{0.8cm}C{1.2cm}C{1.2cm}C{1.2cm}C{1.cm}} 
\hline \hline
 &\textit{a} &\textit{d} &\textit{l}$_1$ &\textit{l}$_2$ \\ \hline  
P &6.43 &1.27 &2.28 &2.24\\         
As &7.08 &1.42 &2.53 &2.49\\        
Sb &8.05 &1.67 &2.91 &2.87\\
Bi &8.47 &1.76 &3.05 &3.02\\ \hline
\end{tabular}
\label{table1}
\end{table}


To check the dynamical stability of these 2D materials, we calculate their phonon spectrums. These OT allotropes have similar phonon dispersions. The phonon spectrum of the OT-Bi bilayer is shown in Fig.~\ref{figure1} (b) as an example. Generally speaking, the phonon bands can be separated into four bunches, and quantitatively, lighter materials have higher phonon frequencies. There are no imaginary phonon modes in all cases, indicating that these 2D allotropes are dynamically stable. Additionally, first-principles molecular dynamics simulations were performed for over 9 ps time, corresponding to 6000 simulation steps. All these allotropes maintain their structures at temperatures up to 600 K, showing good thermal stability.

\begin{figure*}[!htb]
\includegraphics  [width=16cm]{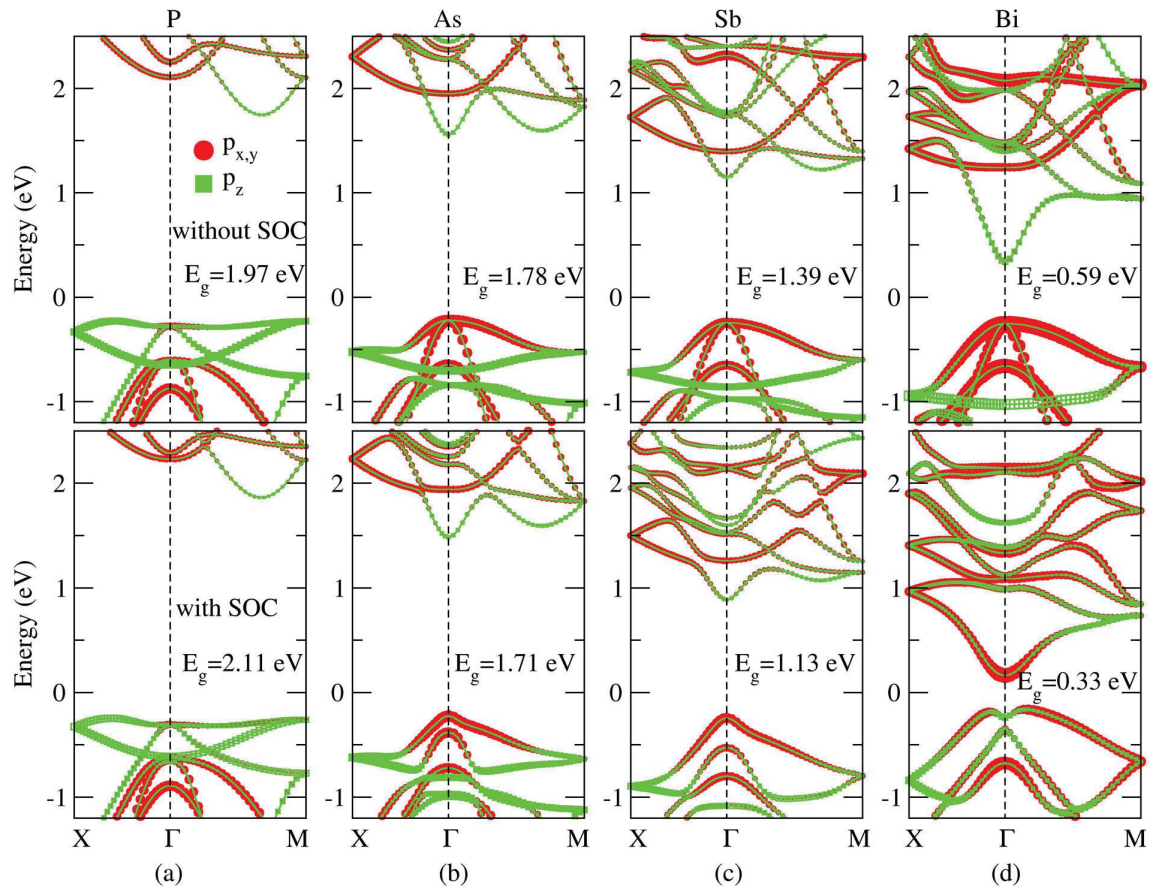}
\caption{Calculated band structures of OT-P (a), OT-As (b), OT-Sb (c) and OT-Bi (d) without (top panels) and with (bottom panels) SOC.  The corresponding global energy gap $E_g$ is also shown. Red and green symbols denote projections of the p$_{x,y}$ and p$_z$ states, respectively, and symbol size indicates the contribution weight.}
\label{figure2}
\end{figure*}

Now we turn to the electronic structures of these 2D allotropes. The calculated energy band structures without and with spin-orbit coupling (SOC) are shown in Fig.~\ref{figure2} for all the 2D OT allotropes. The contributions of the p$_{x,y}$ and p$_z$ orbitals to the wavefunctions are shown in different colors, and the energy gaps are labeled.

\begin{figure}
\includegraphics [width=7.6cm] {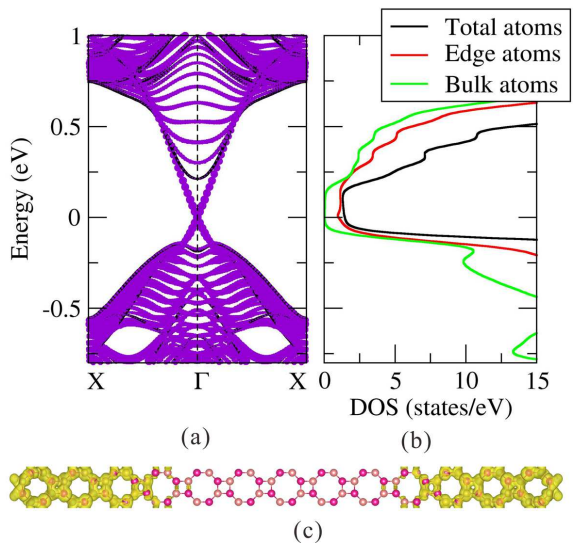}
\caption{Electronic structures of the OT-Bi nanoribbon: (a) Energy band dispersion, (b) projected density of states (PDOS), and (c) real space projection of the Dirac point wave functions. Purple symbols in (a) indicate contributions of edge atoms, and symbol size denotes the weight contribution.}
\label{figure3}
\end{figure}

As shown in Fig.~\ref{figure2}, all the OT bilayers are insulators. OT-P is an indirect gap insulator. Both its conduction band minimum (CBM) and the valence band maximum (VBM) are located away from $\Gamma$ point and they are mainly composed of p$_z$ states. At $\Gamma$ point, the VBM has contributions from both p$_{x,y}$ and p$_z$ states, while the CBM mainly comes from p$_{x,y}$ states. OT-As and OT-Sb allotropes are direct gap insulators. Both their VBM and CBM are located at $\Gamma$ point. The former mainly comes from p$_{x,y}$ states, while the latter mainly comes from p$_z$ states. Besides, from P to Bi, the energy gap decreases monotonically. This is in agreement with the increasing metallic behaviours from P to Bi. The gaps of OT-P, OT-As and OT-Sb range from about 1 to 2 eV, almost covering the full gap range of TMDCs such as MoS$_2$, MoSe$_2$, WS$_2$ and WSe$_2$ \cite{p32}, thus they may be alternative candidates for various near-infrared optoelectronic applications. We note that the energy gaps of OT-As and OT-Sb with SOC are smaller than those without SOC. This can be understood as the following. Spin-orbit coupling has the effect to invert the frontier orbits. If the non-SOC energy gap is not too large and the SOC strength is large enough, band inversion usually occurs, which is the typical mechanism of topological insulators \cite{p20,p21}. The stronger the SOC, the larger the gap opens. Here, As and Sb atoms have weak SOC, and their non-SOC energy gaps are large (all larger than 1.3 eV). Therefore, their SOC is not enough to invert the bands, but only narrowing their gaps. However, the minimum gap (2.11 eV) of OT-P with SOC is slightly larger than the non-SOC gap (1.97 eV), as shown in Fig.~\ref{figure3} (a). This is because of the indirect gap nature in OT-P. Considering the direct gap at $\Gamma$ point, where band inversion usually occurs, the SOC energy gap is 2.38 eV, smaller than the non-SOC gap, which is 2.53 eV.

When not considering SOC, OT-Bi has similar band structures as OT-P and OT-Sb. It has a direct gap, and the p$_{x,y}$ and p$_z$ states contribute to the VBM and CBM, respectively, as shown in Fig.~\ref{figure2} (d). The non-SOC band gap is 0.59 eV, the smallest among the four OT allotropes due to its strongest metallic properties. When SOC is switched on, things become different. The strong SOC of the heavy Bi atoms inverts the VBM and CBM at $\Gamma$, and now the former mainly comes from the p$_z$ states while the latter mainly from p$_{x,y}$ states. The VBM bands around $\Gamma$ resemble an M shape, giving the OT-Bi an indirect energy gap of 0.33 eV, very close to the bulk gap of black P (about 0.3 eV) \cite{p22}. Generally, band inversion is a strong indicator of topologically nontrivial band structures, and we will confirm this in the following.

We now discuss the topological properties of the band structures in these 2D allotropes. To do this, the topological invariant, i.e., the Z$_2$ number is calculated using the parity method \cite{p33}. With this method, the Z$_2$ invariant can be determined through computing the parities of wave functions at time-reversal invariant momenta (TRIMs) in the Brillouin zone as the following \cite{p33},
\begin{equation}
\begin{array}{l}
 \delta_i=\prod_{m=1}^N\xi_{2m}(\Gamma_i),  (-1)^\nu=\prod_i\delta_i.
\end{array}
\end{equation}
Here, $\xi=\pm1$ is the parity eigenvalues of wave functions at four TRIMs, \textit{N} the number of occupied bands, and $\nu$ (mod 2) the Z$_2$ number. If Z$_2$=1, the system is nontrivial, otherwise it is trivial.

The calculated Z$_2$ numbers for the four OT allotropes are listed in Table~\ref{table2}. Obviously, Z$_2$=1 for the OT-Bi allotrope, thus it is a 2D TI, while Z$_2$=0 for OT-P, OT-As and OT-Sb, indicating that they are trivial insulators. This is in agreement with the band structures shown in Fig.~\ref{figure2}, where band inversion only occurs in OT-Bi. The following discussions on OT-Bi are all based on SOC band structure calculations.

\begin{table}[tbp]
\caption{Calculated parities ($\delta$) and Z$_2$ invariants of the four OT allotropes.}
\centering  
\begin{tabular}{L{1.2cm}C{1.cm}C{1.cm}C{1.cm}C{1.cm}}
\hline \hline
 &P &As &Sb &Bi \\ \hline  
$\delta(\Gamma)$ &+ &+ &+ &$-$\\         
3$\delta$(M) &+ &+ &+ &+\\        
$\nu$ &0 &0 &0 &1\\ \hline
\end{tabular}
\label{table2}
\end{table}

As shown in the bottom panel of Fig.~\ref{figure2} (d), OT-Bi has an indirect gap, which usually occurs in most reported  2D TIs  \cite{p11,p15,p34}. The indirect gap here is 0.33 eV (direct gap at $\Gamma$ is 0.40 eV), which is large enough for room temperature observations and utilizations. As far as we know, this gap is the largest among the reported elemental 2D TIs. For elemental TIs without strains or chemical decorations, the largest gap before the present work occurs in the buckled honeycomb Bi(111) bilayer, which is about 0.2 eV \cite{p14,p15}. Gaps of other 2D TIs range from a few meV to 0.1 eV, for examples, 1.55 meV in silicene, 23.9 meV in germanene \cite{p7}, 0.1 eV in tin films \cite{p11}, and 0.1 eV in ZrTe$_5$ and HfTe$_5$ single-layers \cite{p34}. The rather large gap here is due to the strong spin-orbit coupling of the heavy Bi atom. Besides, the interesting octagonal tiling structure here may provide proper Bi-Bi distances suitable for the strong SOC.

2D TIs are bulk insulating, with Dirac-type gapless states at the edges \cite{p20,p21}. To confirm this we construct a nanoribbon from the OT-Bi bilayer and calculate its electronic structures. The band structures of the ribbon are shown in Fig.~\ref{figure3} (a), where gapless edge states clearly appear. The Fermi velocity at the Dirac point is 6.18$\times10^5$ m/s, which is larger than the 4.4$\times10^5$ m/s in stanene but it is similar to the 6.8$\times10^5$ m/s in fluorinated stanene \cite{p11}. Fig.~\ref{figure3} (b) shows the projected density of states (PDOS) on both the edge and the bulk atoms. As shown in the figure, edge atoms contribute almost all the DOS around the Fermi level, indicating that the edge is metallic. On the other hand, bulk atoms have negligible contributions near the Fermi level, consistent with the insulating bulk states. Real space projection of wave functions of Dirac states at $\Gamma$ point (Dirac point) in Fig.~\ref{figure3} (c) shows that the edge states here have a large penetration depth of about 3 nm. This is larger than that in the buckled honeycomb Bi(111) bilayer, where the depth is less than 2 nm \cite{p18}. Further analysis indicates that spins and momenta at the Dirac states are always perpendicularly locked. These features of the ultrathin OT-Bi film are consistent with their topologically nontrivial properties \cite{p20,p21}.

\begin{figure*}
\includegraphics [width=12cm]{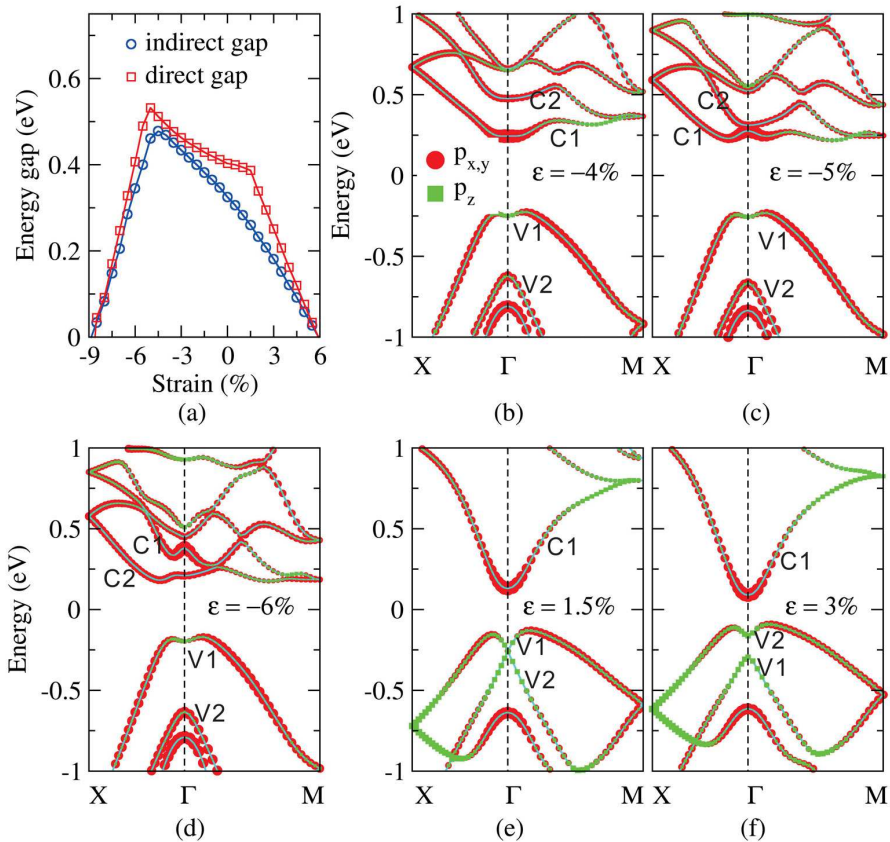}
\caption{ Evolutions of (a) energy gaps and (b)-(f) band structures of the OT-Bi ultra-thin film under different external strains. }
\label{figure4}
\end{figure*}

Nontrivial TIs are robust, i.e., they are able to keep their nontrivial properties against nonmagnetic perturbations, such as strain. Strain is also a widely-used method to improve or manipulate the band gaps of 2D TIs \cite{p7,p11,p34}. In this work, we apply the following external strain
\begin{equation}
\begin{array}{l}
\varepsilon=\frac{a-a_0}{a_0}\times100\%,
\end{array}
\end{equation}
where \textit{a}$_0$ is the equilibrium lattice constant, and \textit{a} is the strained lattice constant. So $\varepsilon>0$ ($<0$) means tensile (compressive) strains. Fig.~\ref{figure4} (a) shows the band gap of the OT-Bi bilayer as the function of the applied strain $\varepsilon$. The energy gap remains open for strains between $\varepsilon=-9\%$ and 6\%, indicating that no topological phase transition occurs within this range.

As shown in Fig.~\ref{figure4} (a), the gap value is very sensitive to strain. We can understand this by analyzing the evolution of the four frontier bands, C1, C2 and V1, V2 as indicated in Fig.~\ref{figure4} (b) to (f). With compressive strains, V1 and V2 bands change very slowly, C1 and C2 bands shift significantly. The C1 band shifts upward, while the C2 band shifts downward, so the energy gaps increase at the initial stage of compression. At about $\varepsilon=-5\%$, the C1 and C2 bands touch each other at $\Gamma$ point, where the direct and indirect gaps reach their maximum values of 0.53 and 0.48 eV, respectively. Under further compressive strains, the C2 band becomes the lowest conduction band and shifts down, and the energy gaps decrease with increasing compressive strain. Under tensile strain, the C1 and C2 bands change very slowly, while the V1 and V2 bands shift upward. So the energy gaps decrease with tensile strain. The V2 band moves faster than the V1 band, and they touch at $\Gamma$ point when $\varepsilon=1.5\%$. Under further tensile strain, the two bands exchange order around $\Gamma$, now the direct gap at $\Gamma$ point is between the C1 and V2 bands. So the direct gap decreases at a faster rate with tensile strain beyond 1.5\%. The indirect gap is still determined by the C1 and V1 bands, because the band exchange only occurs in a small area around the $\Gamma$ point of the Brillouin zone. In this strain process, the energy gap never reaches zero, so the Z$_2$ number of the system does not change. Strain is an often used method for manipulating the gaps of 2D TIs. For example, it has been reported that a 20\% tensile strain can increase the energy gap of the single-layer ZrTe$_5$ from 0.1 eV to more than 0.2 eV \cite{p34}, and a compressive strain of $-6\%$ can change the gap of silicene from 1.55 meV to 2.90 meV \cite{p7}.

In addition, the OT-Bi quadruple layer is also calculated to be a 2D TI with an energy gap of 0.15 eV (the direct gap is 0.37 eV). Films with even more layers become metallic. It is possible to modify their electronic properties by methods such as surface adsorptions \cite{p16}, which is of interest for future studies.

\vspace{5mm}
\textbf{Discussion}

Different crystal structures of the same element can possess different physical properties. For example, in contrast to the OT-P bilayer studied here, the black phosphorus bilayer in puckered honeycomb structure is a direct gap insulator \cite{p24}. Its VBM and CBM are both located at $\Gamma$. The OT-P bilayer has a larger band gap than the BP bilayer, which was calculated to be about 1.51 eV. For Bi, both the OT-Bi bilayer and the Bi(111) bilayer are indirect gap insulators \cite{p14}. However, the energy gap of the OT-Bi bilayer is also larger than the gap of the Bi(111) bilayer in distorted honeycomb structure.

These OT allotropes have the square lattices. For experimental synthesis on a substrate, candidate substrates are zincblende (ZB) CdTe, InSb, and HgTe (for OT-P), ZB-BN, BAs (for OT-As), Ge, ZB-AlAs, GaAs (for OT-Sb), and ZB-CdSe, GaSb, InAs, ZnTe (for OT-Bi). All these substrates can provide square lattices with lattice mismatch less than 2\% for the corresponding 2D octagonal tiling allotropes. The experimental fabrications of these OT allotropes, the band structure engineering by methods such as surface adsorption, and properties of thicker films are all of great interests, and further theoretical and experimental explorations along these directions are needed.

In conclusion, we predict a new structure of 2D allotropes of group-\uppercase\expandafter{\romannumeral5} elements in octagonal tiling structures. They are insulators with energy gaps ranging from about 0.3 to 2.0 eV, covering almost the whole gap range of BP and TMDCs, so they may be promising candidates for various near and mid-infrared optoelectronic applications, such as photodetectors and modulators. In particular, the octagonal tiling Bi allotrope is predicted to be a 2D topological insulator, and its energy gap is larger than any previously reported elemental 2D TIs. We hope that the present findings will motivate further experimental and theoretical efforts.

\vspace{5mm}
\textbf{Methods}

This work is performed using first-principles method within the framework of density functional theory of generalized gradient approximations (DFT-GGA) \cite{p35,p36}, as implemented in the VASP codes \cite{p37}. Phonon spectrum is calculated using the phonopy package \cite{p38}. A vacuum of more than 20 \r{A} is added between the layers in the supercell calculations of the 2D bulks or 1D ribbons. Both the in-plane lattice constants and the atom positions of 2D bulks are relaxed until the force on each atom is less than 0.01 eV/\r{A}. A 17x17 K-point mesh is used for all the relaxation and static calculations, and spin-orbit coupling is explicitly included.

\vspace{5mm}
\textbf{Acknowledgements}

This work was supported by National Natural Science Foundation of China (Grant Nos.\ 11474197 and 11521404) and National Basic Research Program of China (Grant No.\ 2013CB921902). P. Li acknowledges the support of Natural Science Fundation of Anhui Province (Grant No.\ 1308085QA05). Computations were performed at the HPC of Shanghai Jiao Tong University.

\end{document}